\documentclass[reprint,aip,amsmath,amssymb,nofootinbib]{revtex4-2}
\usepackage{amssymb}
\usepackage{lipsum}
\usepackage{amssymb}
\usepackage{bbold}
\usepackage{amssymb}
\usepackage{graphicx}
\usepackage{graphics}
\usepackage{amsmath}
\usepackage{placeins}
\usepackage{hyperref}
\usepackage[dvipsnames]{xcolor}
\usepackage{wasysym}
\usepackage{soul}
\usepackage{float}
\usepackage{array}
\usepackage{tabulary}
\usepackage{caption}
\usepackage{fancyhdr}
\usepackage{wrapfig}

\usepackage[english,ngerman]{babel}

\usepackage[normalem]{ulem}
\usepackage{graphicx}
\usepackage{color,soul}
\usepackage{subfigure}
\usepackage{dcolumn}
\usepackage{bm}
\usepackage{hyperref}
\usepackage{url}
\usepackage{multirow}
\hypersetup{
    colorlinks=true,
    linkcolor=blue,
    filecolor=blue,      
    urlcolor=blue,
    citecolor=blue,
}

\usepackage[makeroom]{cancel}

\begin{document}
\selectlanguage{english}
%\title{Fabrication and Characterization of Dual-mode 3D Printed RF Cavities }
\title{3D Printing as a Rapid Prototyping Approach for Novel RF Cavity Designs}

    \author{\firstname{David} \surname{Sims}}
    \email{simsdav4@msu.edu}
    \affiliation{Department of Electrical and Computer Engineering, Michigan State University, MI 48824, USA}
    \affiliation{Department of Physics and Astronomy, Michigan State University, East Lansing, MI 48824, USA}

    \author{\firstname{Benjamin} \surname{Sims}}
    \email{simsben1@msu.edu}
    \affiliation{Department of Electrical and Computer Engineering, Michigan State University, MI 48824, USA}
    \affiliation{Department of Physics and Astronomy, Michigan State University, East Lansing, MI 48824, USA}
    \affiliation{Facility for Rare Isotope Beams, Michigan State University, East Lansing, MI 48824, USA}

        \author{\firstname{Brian} \surname{Wright}}
	\affiliation{Department of Electrical and Computer Engineering, Michigan State University, MI 48824, USA}
    
    \author{\firstname{John W.} \surname{Lewellen}}
    %\email{jwlewellen@lanl.gov}
    \affiliation{Accelerator Operations and Technology Division, Los Alamos National Laboratory, NM 87545, USA}

        \author{\firstname{Sergey V.} \surname{Baryshev}}
    \email{serbar@msu.edu}
	\affiliation{Department of Electrical and Computer Engineering, Michigan State University, MI 48824, USA}
	\affiliation{Department of Chemical Engineering and Material Science, Michigan State University, MI 48824, USA}

\begin{abstract}
   %This research investigates the fabrication, and characterization of 3D-printed tabletop cavities for electromagnetic and rf applications relative to a characterized full copper parent cavity of identical dimensions. 
   3D-printing of radiofrequency (RF) cavity resonators could provide a cost-effective solution that enables rapid prototyping and design flexibility compared to traditional fabrication of full-metal cavities. In this work, the feasibility of fabrication of a useful multi-mode GHz cavity is explored. Two kinds of plastics, two slicing approaches and two metal coating techniques were used to build a series of clamped cavities with thin inner copper surface on otherwise 3D printed plastic surface. The cavities were then bench-tested to identify spatial field distributions, operating frequencies and quality factors ($Q$-factor). Pros and cons of the used fabrication approaches were identified and understood, and the performance of longitudinally sliced painted cavity design demonstrated considerable practicality of 3D-printing approach in designing rf systems.

   %When directly compared to a full copper version and the 3D printed variant, including cavity performance, field strength, and $Q$-factor performance.
   %These metrics provide a important metric for knowing the relative performance of full metal cavity given a 3D printed prototype. This research evaluates the feasibility of 3D-printed cavities in RF applications in the prototyping stage and aims to highlight areas for optimization in future research.

\end{abstract}

\maketitle

\section{Introduction}\label{intro}

Traditional methods of radiofrequency (RF) cavity fabrication involve the use of high quality solid conductive materials such as oxygen free copper (OFC), to be machined into precise geometries to meet the requirements of their respective high-frequency applications. When bench-testing novel geometries or designs, traditional fabrication methods using OFC materials can quickly become expensive in both time and cost, caused by raw material supply chain and fabrication lead time. Contemporary additive manufacturing, like 3D printing, is capable of producing geometrically complex and accurate designs. Hence, 3D printing could become an alternative fabrication approach offering fast and inexpensive means of experimentally testing and verifying computational cavity designs.

It must be noted that in the field of electron paramagnetic resonance TE$_{102}$ cavities were previously manufactured as metal-plated glass and plastic structures as early as in the 1950's \cite{glass-cavity-1958, plastic-cavity-1959, plastic-cavity-1969}. While those cavities sufficed EPR application, the earlier prototypes were rather simplistic and, by literature review, did not go through basic microwave characterization established around the same time \cite{ginzton_microwave_1957} where, at the very least, resonance frequency, quality factor, and field distribution need to be measured. More recently, additive manufacturing has emerged as a cost-effective method for measuring complex permittivity of technologically important liquids such as oils \cite{mohammed_3d_2021,rocco_span_2021}. Various clamp-type cavities were reported where the $Q$-factor varied anywhere from $\sim$300 to 6,000. However, the reasons behind such a large variation were not given in the corresponding literature.

%By fabricating cavities capable of sustaining high quality factors, a dielectric sample can be inserted, and the resulting perturbation measured with a VNA. From the observed frequency shift, the material's permittivity can be extracted \cite{mohammed_3d_2021,rocco_span_2021}. This method requires achieving and maintaining a high quality factor to ensure accurate frequency tracking. Although these previous experiments have gone into great detail regarding the quality factor testing, other critical parameters regarding field distribution have not been explored thoroughly. Further understanding of the characteristics of the cavity is necessary for optimization. As with novel cavity geometries, significant performance may be lost within the fabrication process if decisions are not made congruent to ensuring the inner shell of the cavity is as seamless as possible. This requires fabricating multiple cavities of identical geometry and performing congruent analyses to evaluate the impact of the manufacturing process on performance.

This study fills these previous gaps and explores the feasibility of a 3D printing approach to fabricate a workable and high-complexity RF cavity that has a novel dual-mode design. Few cavities were printed from two different plastics and copper-coated by two different methods. All cavities were successfully bench-tested and demonstrated, though certain important differences were observed. Cavities final performance were quantitatively compared in terms of electric field distributions and $Q$-factors and discussed in detail.

%The copper parent cavity is a dual mode cavity which offers a unique comparative advantage. As it is designed to accommodate two modes at different frequencies we are able to test the affects of different frequencies within our 3D printed model. Additionally showing that a fast and inexpensive 3D printed cavity is able to accommodate both modes is impactful for the validity of testing non traditional designs in this medium.
%By investigating different fabrication methods, such as sputtering and painting, and analyzing the impact of slicing direction during printing, this research seeks to provide insights into the feasibility of using 3D-printed cavities for high-frequency applications.

\section{The Concept of Advanced Cavity under Study}\label{Printing process}

In modern high-brightness RF guns, ultimate attainable electron beam spatio-temporal coherence is constrained between two non-linear forces, i.e. space charge and RF synchronization \cite{kim1989rf}. To address this basic issue, Serafini $et$ $al.$ \cite{serafini1992} proposed a multi-mode RF gun that would serve to linearize phase (coordinate-momentum) space of the accelerated electrons.
\begin{figure}
	\includegraphics[width=8cm]{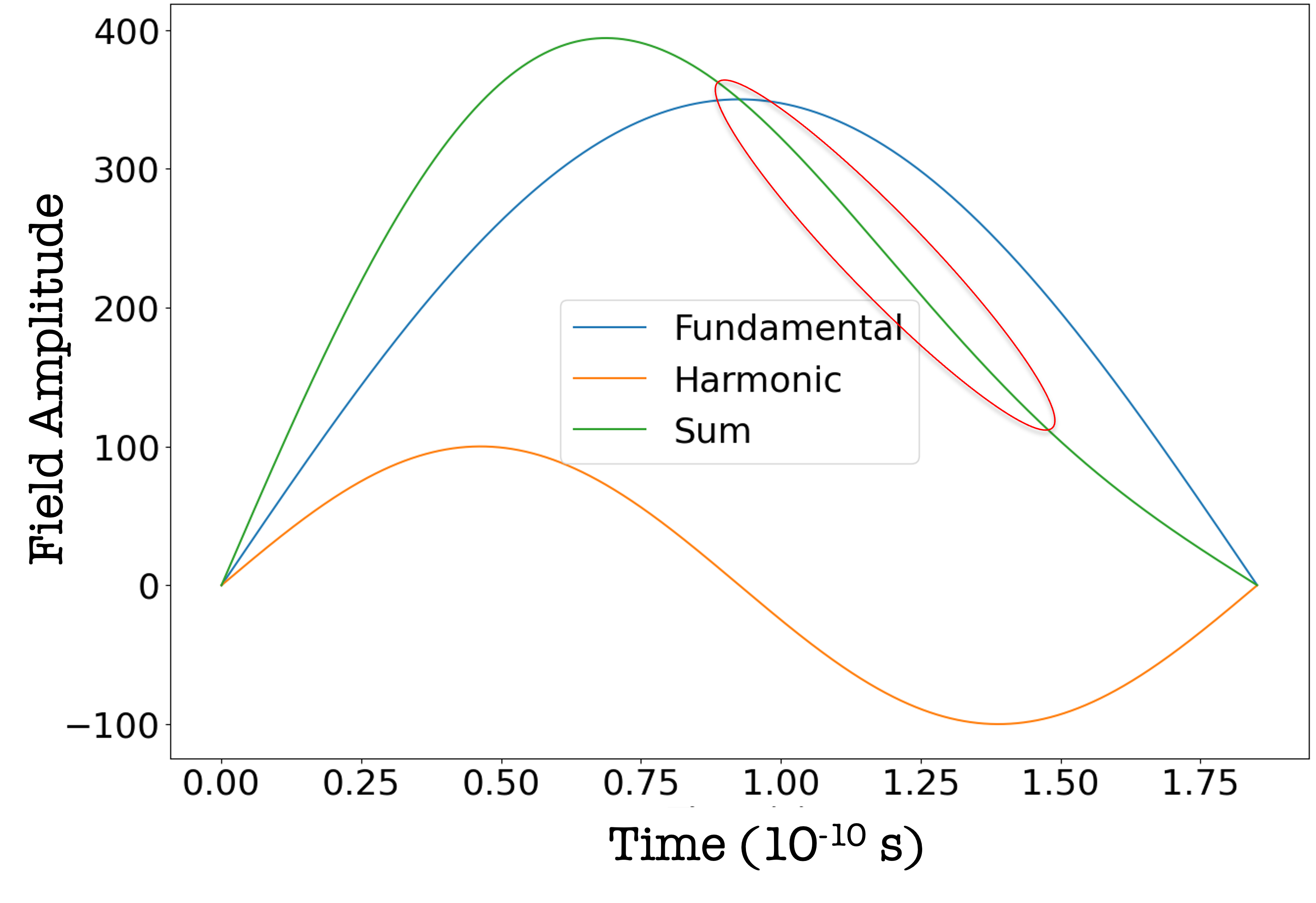}
	\caption{The concept of TM$_{010}$ (blue) and TM$_{011}$ (orange) superposition highlighting the linearization of the $E$-field.} 
	\label{dual}
\end{figure}
Various computational studies \cite{dowell_two-frequency_2004, lewellen_higher-order_2001, lewellen_field-emission_2005} proved that multi-mode cavities could potentially reshape and advance the high power RF technology. However, such designs were never realized and vetted in practice, because building and tuning such cavities are practically challenging and require significant time and resources if manufacturing is solely based on OFC technology.

In this work, we consider a cavity that is required to host and individually tune a combination of fundamental TM$_{010}$ 3 GHz and higher-order mode TM$_{011}$ 6 GHz. As schematically illustrated in Fig.~\ref{dual}, individually manipulating with the amplitude and phase of the two modes would allow to linearize the superposition electric field (shown by red ellipse) hence allowing for linear energy chirp applied to electrons. Such TM$_{010}$/TM$_{011}$ cavity would be a beam compression device that does not sacrifice coherence of the beam as compared to the traditional approach used in so-called bunching RF cavities \cite{trad-buncher}.

\section{Design and fabrication}\label{Printing process}

\subsection{Plastics and Printing}

The dual-mode cavity was fully designed in COMSOL which included all ports for couplers and tuners. A Stratasys J55 Prime 3D machine was used to print multiple latitudinally and longitudinally sliced tests cavities.
%Each 3D printed cavity underwent a printing process that took 7 hours and 35 minutes, utilizing 
VeroUltraWhite RGD824, a photopolymer material, was used as a feedstock material for printing. After printing, the cavities were cured with UV light inside the printer to ensure the material hardened appropriately. %The final weight of each print was recorded as XXXXX grams.

\begin{figure}
        \includegraphics[width= 8 cm]{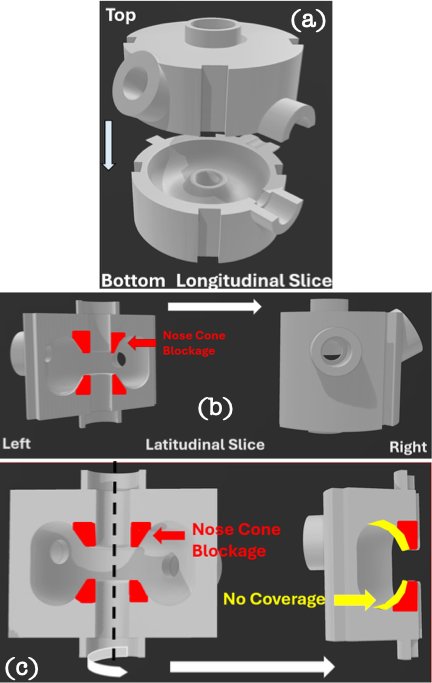}
	\caption{3D models of two distinct cavity slicing: (a) longitudinal and (b) latitudinal where nose cones blocking sputtering are shown in red. (c) Another rendering of blockade locations due to the complex nose cone geometry.}
	\label{slices top and bot}
 
\end{figure}

Because interior conductive layer had to be applied on the inner cavity surfaces, to store the electromagnetic energy,
%thus allowing for a standing wave to form.The cavity's interior design
required that the cavity be printed in two halves.
%This was necessary to facilitate the coating of the internal surface with a conductive layer.
Two possible options, longitudinal and latitudinal slicing (Fig.~\ref{slices top and bot}), were explored. This change in slicing direction aimed to provide insight into how the orientation of the print might influence the overall functionality, which will be explored in greater detail in Section \ref{Discussion}.
%axes for each cavity, This can be seen in figure \ref{slices top and bot}. The geometry of the nose cones based upon the geometry of the slice of the cavity acted, particularly the longitudinally sliced cavity, obstructed the even coating during the sputtering process. 

Once printed, the cavities were subjected to two distinct methods of coating the interior with a conductive material. The first method involved sputtering, while the second method used a conductive paint.

\begin{figure}
	\includegraphics[width=9cm]{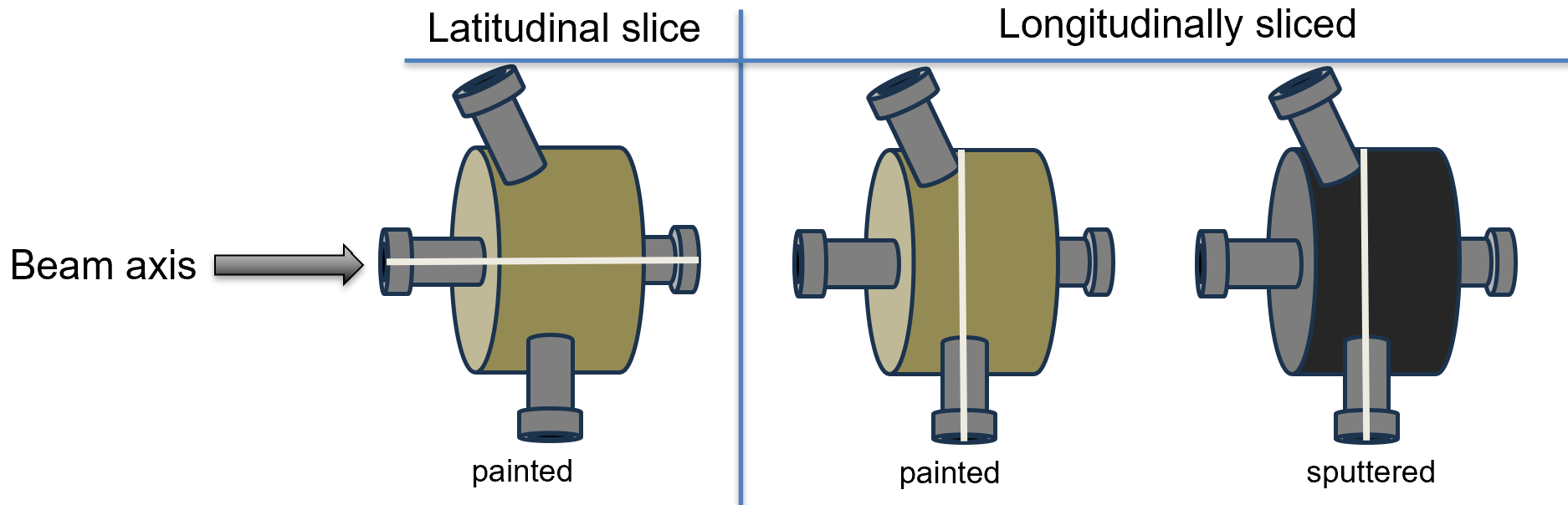}
	\caption{Diagram of the three assembled and tested cavities.}
	\label{Cavity Design Table}
 
\end{figure} 

As presented in Fig.~\ref{Cavity Design Table}, a total of 3 cavities were successfully fabricated. A fourth cavity was attempted, which utilized a sputtering process on a cavity sliced along the longitudinal axis. However, due to the complex geometry of this cavity, the copper coating was uneven, leading to suboptimal and nonviable results. Namely, the geometry of the nose cones, based on the geometry of the slice of the cavity (shown in red in Fig.~\ref{slices top and bot}), creates pockets with no direct line of sight (shown in yellow), thereby obstructing the sputtering process. 
%This issue highlights the challenges associated with the geometry of 3D printed parts when subjected to certain manufacturing processes. Further investigation will be needed to optimize the sputtering process for such geometries in future studies.

\subsection{Sputtering Process}
The cavities were first cleaned using isopropanol to ensure a grease-free surface. It was then placed into a Denton Vacuum Desktop Pro DC dual-gun sputtering system. The system was pumped down to a base pressure of $1\times 10^{-5}$ Torr.
%, a process that took a couple of hours to stabilize. Once the desired vacuum was achieved, high-purity argon gas was introduced into the chamber for purging, typically performed twice to ensure a clean environment.
Using argon as a working gas, first, a 30 nm layer of titanium was deposited onto the cavity to improve adhesion of the copper layer. Using a calibration curve with a particular ion energy and current, a total of 5 $\mu$m of copper had been sputtered. This method ensured that the cavity's conductive coating was much thicker that the skin depth, about 1 $\mu$m at 3 and 6 GHz.
%, providing the necessary skin depth for optimal performance. The skin depth $\delta$ is defined as:
%\[
%\delta = \sqrt{\frac{2}{\mu \sigma \omega_0}},
%\label{eq:1}
%\]
%Here $\mu$ is the permeability of the material for copper, $\sigma$ is its electrical conductivity, and $\omega_0$ is the angular frequency of the oscillating field. In the case of copper at the fundamental mode (i.e 2.81 Ghz) the skin depth is around 3.12 microns. The copper depth of 5 microns allows for tolerance of the system and approximately twice the amount of skin depth. 

\subsection{Painting Process}
The cavities were first cleaned using  isopropanol to remove contaminants that could interfere with the painting process. This ensured for more optimal adhesion of the paint to the plastic surface. Once cleaned, the cavity was then dried for 24 hours and prepared for the application of the 842WB Super Shield Water-Based Silver Conductive paint. The first layer of paint was then applied to the inner surface of the cavity using a fine brush, which ensured controlled application. Efforts were made to keep the applied layer as uniform as possible by visual inspection.
%Any significant variations in the paint thickness and uniformity would unintentionally alter the cavity's dimensions, impacting the wall loss that occurs within the cavity and mode structure. 

This process was repeated three times, each layer was allowed to dry for 30 minutes before the next coat was applied. After the third layer had dried, a final layer of paint was applied around the edges of the cavity to ensure a complete and uniform conductive surface. Following the final application, the cavity was then closed and sealed by joining the two halves and tightening them together using custom clamps. This ensured a secure seal between the two halves. The seal was left to dry for two days to fully cure and bond the two halves of the cavity together. This curing time ensured stability of the conductive layer and the overall integrity of the cavity prior to any testing or further processing.

%\begin{figure}
%	\includegraphics[width=7cm]{Latitudinal Slice paint.png}
%   \includegraphics[width=7cm]{Longitudinal slice paint.png}
%	\caption{ Paint locations for the latitudinally and longitudinally sliced cavities (points of contact marked in orange) }
%	\label{paint}
 
%\end{figure} 

\section{Characterization results}\label{Characterization}

\subsection{Wire Pull}

\begin{figure}
	\includegraphics[width=7cm]{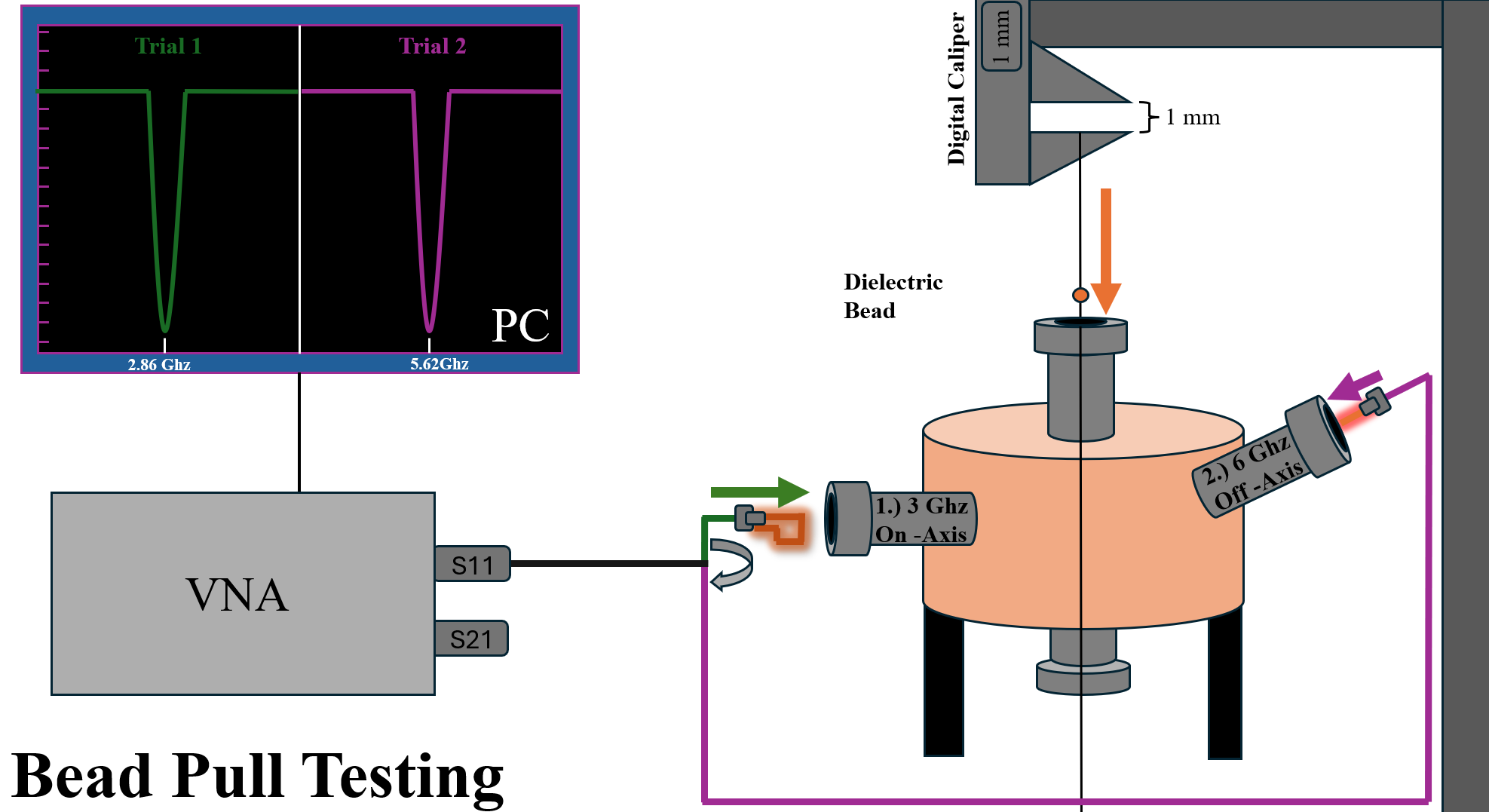 }
	\caption{Conceptual diagram of the bead-pull testing setup/procedure.}
	\label{BeadPull}
 
\end{figure}

In order to measure the mode $E$-field distributions, localized perturbations were introduced to the electromagnetic field using a short, thin piece of wire attached to a dielectric string and pulled through the cavity along its beam axis. As shown in the diagram in Fig.\ref{BeadPull}, more specifically, a 16-gauge wire, 3 mm in length, was attached to a dielectric string and pulled through the longitudinal axis of the cavity in 1 cm ± 10 mm intervals across all three fabricated 3D-printed cavities. Identical type of measurements were made for both TM$_{010}$ and TM$_{011}$ modes. At every position, frequency detuning $\Delta\omega_{dt}(z)$ along the beam/longitudinal/$z$-direction was recorded: the resonant frequency of the cavity mode shifts in proportion to the strength of the $E$-field (on-axis, $z$-axis, TM mode $H$-field is negligible) as 

\begin{gather}
    \Delta\omega_{dt} (z)= \omega(z) - \omega_0 \propto E(z)^2 \omega_0, \label{eq:2} 
\end{gather}

\begin{figure}
	\includegraphics[width=9cm]{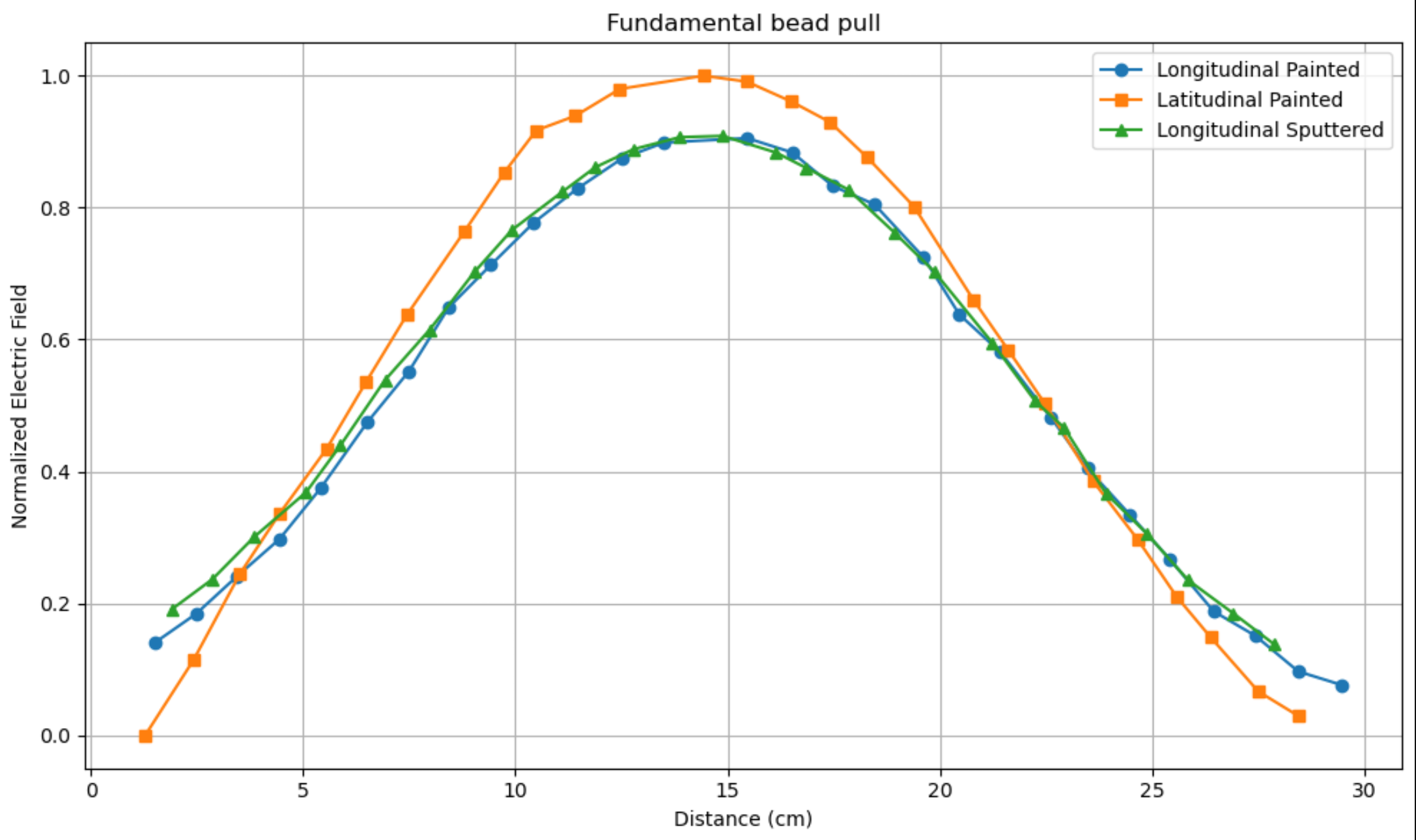}
	\caption{Relative $E$-field for TM$_{010}$ mode.}
	\label{Fundamentalbeadpullresult}

	\includegraphics[width=9cm]{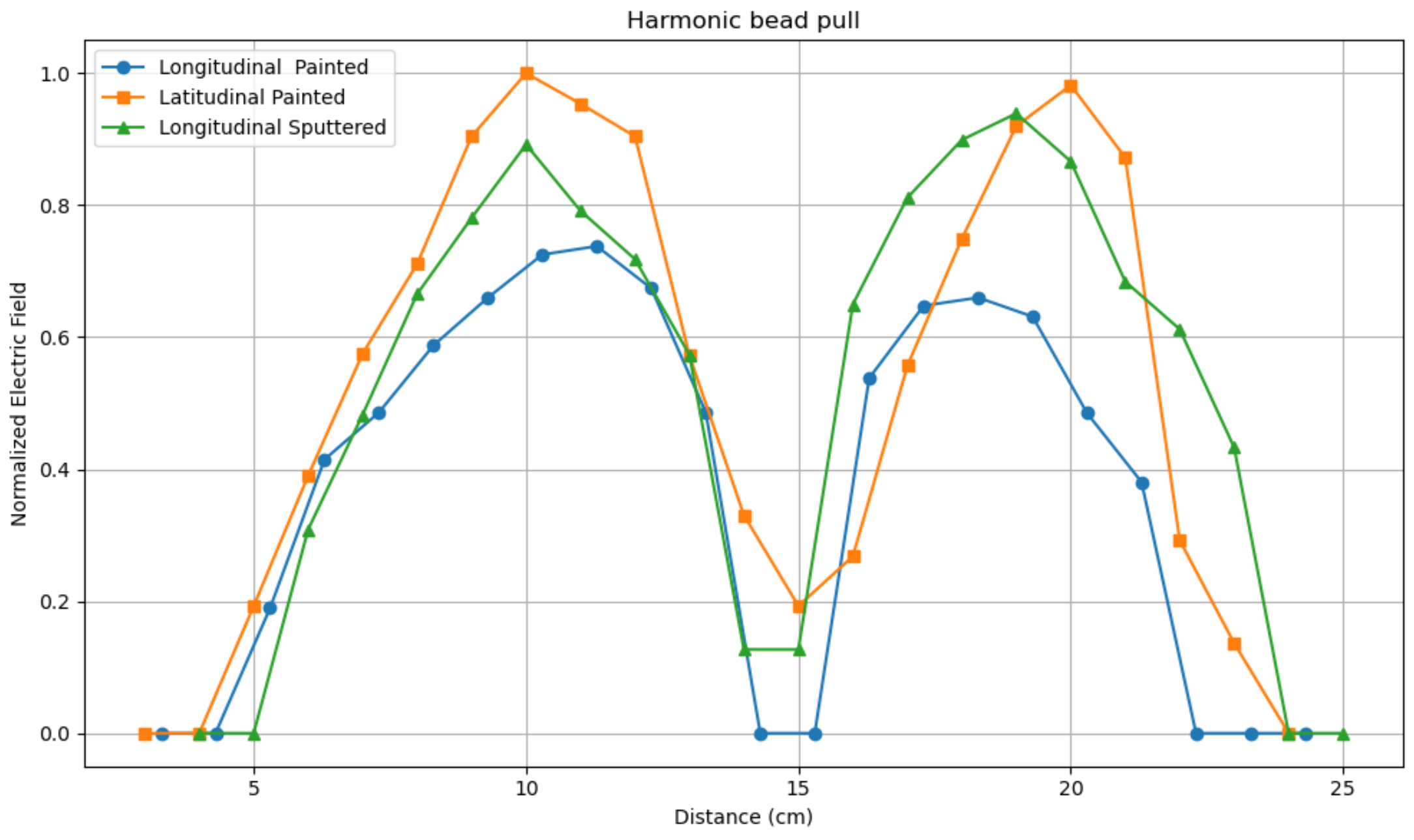}
	\caption{Relative $E$-field for TM$_{011}$ mode.}
	\label{Harmonicbeadpullresult}
 
\end{figure}

The wire-pulls successfully demonstrated that the designed and fabricated cavities indeed were able to store both prescribed modes, making them dual-mode cavities. TM$_{010}$ and TM$_{011}$ modes were successfully excited as can be seen in Fig.\ref{Fundamentalbeadpullresult} (fundamental mode) and Fig.\ref{Harmonicbeadpullresult} (harmonic).
%However, while the mode structure was preserved, the magnitude of the relative \(E\)-field intensity was consistently lower than that in the full copper parent cavity, leading to a reduction in the efficiency of mode excitation and wave propagation.

\subsection{Quality Factor}

\begin{figure}
	\includegraphics[width=7cm]{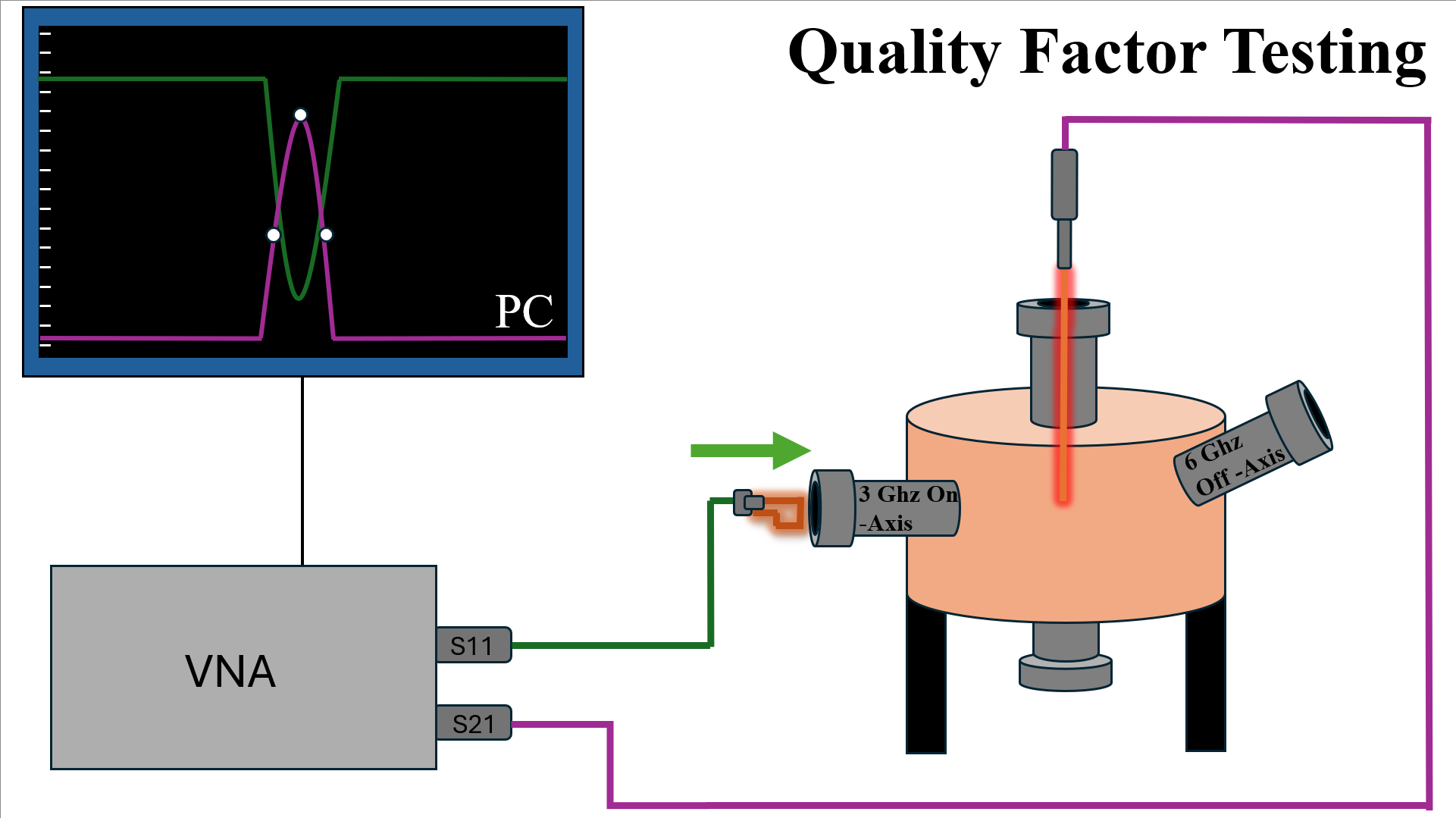}
	\caption{Conceptual diagram of the $Q$-factor testing setup/procedure.}
	\label{Q factor testing}
\end{figure}

To estimate mode $Q$-factors S11 and S21 parameters were measured using a calibrated vector network analyzer (VNA). To ensure minimal power loss between the input transmission line and the cavity, impedance matching was performed using a Smith chart. The input probe (S11) was adjusted, and the impedance of the system was plotted on the chart. By manipulating the location of the input probe within the cavity, the impedance point of the cavity was aligned with the 50 $\Omega$ reference point at the center of the Smith chart. This alignment ensured that the cavity's impedance matched that of the input transmission line, minimizing the reflection coefficient and allowing for maximum energy transfer. A phase diagram was used to then verify critical coupling.
%by analyzing the phase response of the system across different frequencies.
As the coupler was adjusted, the phase change as a function of frequency of the transmitted signal was observed. Around the resonant frequency, the phase shift exhibited a 180$^{\circ}$ change, highlighting the balance between the electric and magnetic fields of the modes thereby additionally confirming critical coupling. At critical coupling, to evaluate $Q$-factor the pickup probe (measuring S21 parameter) was inserted to reach into the beam axis of the cavity, as diagrammed in Fig.\ref{Q factor testing}. Again, the S21 probe length was optimized to minimize its coupling and therefore cavity perturbance/loading by iteratively changing the probe length such that the width $\Delta \omega$, measured -3dB on either side of the $\omega_0$ of the S21 peak was minimal. After that, so-called loaded $Q$-factor ($Q_L$) was calculated as
\[
Q_L = \frac{\omega_0}{\Delta \omega}.  \label{eq:3}
\]
%along with the frequency -3db on either side. Then, the pickup probe would be removed and reduced by 1 mm in length. This procedure was followed until the pickup probe began to decouple from the cavity. The results, plotted in Figure XX, show the Q-factor measurements corresponding to different probe lengths. As the probe was shortened, the Q-factor increased, indicating a stronger coupling to the cavity at certain probe lengths. The peak of the Q-factor signifies the point at which the pickup probe is weakly coupled to the cavity, indicating that the coupling is optimized for energy transfer without excessive perturbation. At this point, the system achieves the desired minimal power loss, as evidenced by the sharp Q-factor peak.
%\section{Results}\label{Results}
%\subsection{Spatial field }
%\subsection{Measured Q-factor }
%where $\omega_0$ represents the resonant frequency, corresponding to the peak of the S21 response, and $\Delta \omega$ is the bandwidth, measured at -3 dB below this peak.
At minimized loading (S21) and critical coupling (S11),
%the minimal coupling point, the $\beta$ factor approaches 0, indicating that the reflected power from the cavity is negligible. Consequently, the input probe returns to a
the coupling factor $\beta$ is approximately equal to 1. Hence, once $Q_L$ was determined, the intrinsic quality factor ($Q_0$) was calculated as
\[
Q_0 = (1 + \beta)Q_L.
\]
All computed $Q_0$ factors for all three cavities for the both modes are summarized in Table~\ref{T:1}.
%The critical coupling points identified using the methods above yielded the top Q factor for each cavity, as summarized in the table \ref{T:1}:

\begin{table}[h!]
\centering
\caption{Comparison of $Q_0$ for each 3D-printed cavity.}
\begin{tabular}{c|c|c|c}
%\hline
\textbf{Slicing } & \multicolumn{2}{c|}{\textbf{longitudinal}} & \textbf{latitudinal} \\ 
\hline
Coating & painted & sputtered & painted \\ 
\hline
$Q_0^{TM010}$ & 320 & 623 & 1364 \\ 
\hline
$Q_0^{TM011}$ & 336 & 698 & 1423 \\
\end{tabular}
\label{T:1}
\end{table}

\section{Discussion}\label{Discussion}

%To reduce fabrication costs and enable rapid prototyping, 3D-printed cavities were employed to investigate dual-mode RF structures. These polymer-based cavities were coated with conductive layers and evaluated through wire pull measurements to confirm support for both the fundamental and harmonic modes. Quality factor (\(Q\)) measurements were used to quantify performance, with specific emphasis on optimizing slice orientation and plating method. Optimization efforts focused on 3D-printed variants due to their drastically lower mass compared to solid copper, making them attractive for weight-constrained environments.

Summarizing, among the tested configurations, two relevant trends emerged:

\noindent 1) For the longitudinal case, the sputtered copper coatings yielded superior conductivity and higher \(Q\)-factors relative to silver-painted surfaces. This is despite silver having  better conductivity than copper;

\noindent 2) Latitudinal slicing produced $Q$-factors up to four times higher than longitudinal slicing. Due to geometrical shadowing, testing of a latitudinally sputtered cavity was not feasible, because it was not possible to deposit copper into the ballon-shaped areas of the cavity (see Fig.~\ref{slices top and bot}).

\noindent Nevertheless, extrapolating from the current trend suggests that latitudinal slicing with sputtered copper would yield the highest $Q$-factor among those studied. Let us now discuss in more detail about how and why material and slicing factors affect $Q$-factor metrics.

%\begin{figure}
%	\includegraphics[width=9cm]{Impedence ratio.png}
%	\caption{Depicts the ratio of Impedance to the coupling of high and low fields.}
%	\label{Impedence ratio}
 
%\end{figure} 

\begin{figure}
	\includegraphics[width=7cm]{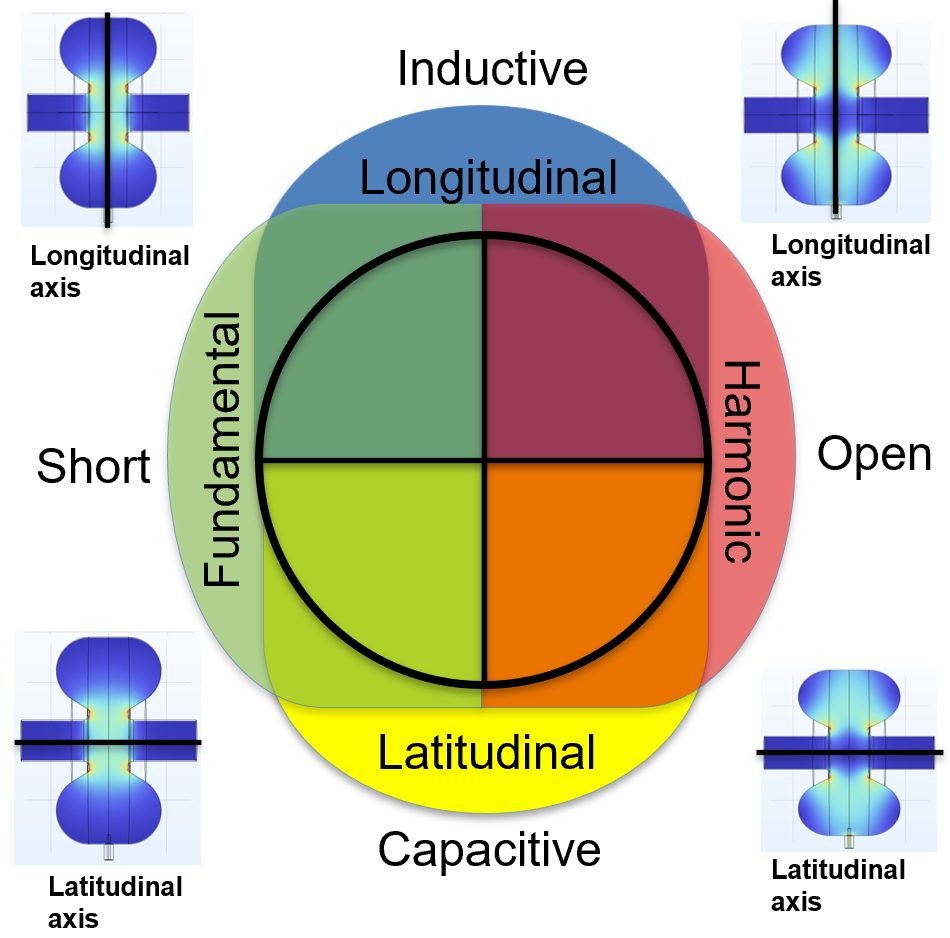}
	\caption{The four quadrants of the Smith chart related to the electric field for both the fundamental and harmonic modes.}
	\label{Smith Chart}
 
\end{figure}

\begin{figure}
	\includegraphics[width=7cm]{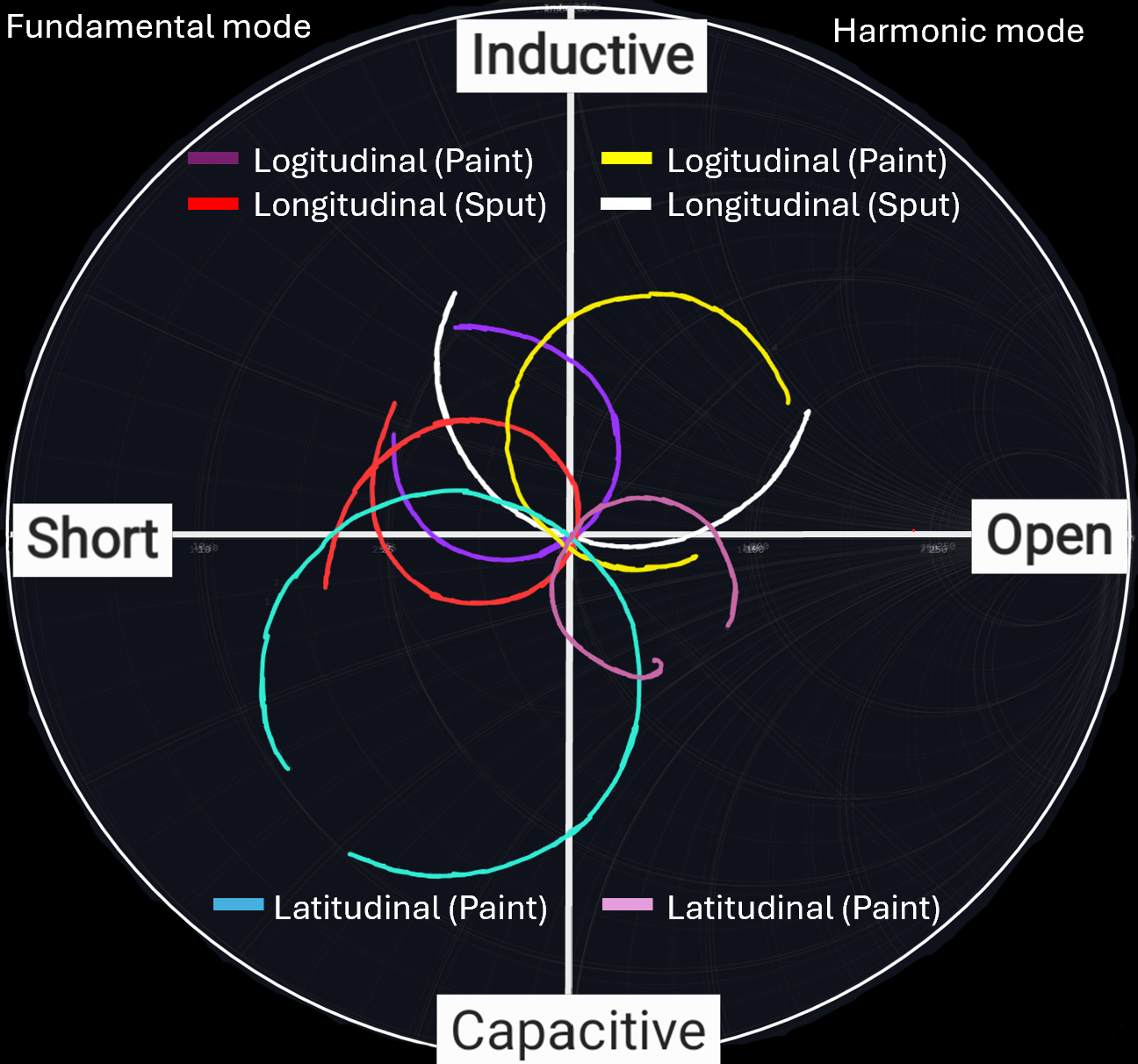}
	\caption{Experimental Smith chart data enhanced for clarity}
	\label{experimental smith}
 
\end{figure} 

\subsection{Geometric Slicing Factor}

Enhancement or degradation in cavity performance largely depends on surface current behavior, which in turn is dependent on
%RF surface currents are confined within the skin depth and flow tangential to the magnetic field. The
slicing. The slice direction introduces interruptions along the cavity surface, which result in significant losses.
%Geometrically, the cavity slice direction alters the surface current breakage, impacting the coupling behavior and overall performance of the resonator.
In RF cavities, surface currents \(\vec{J}_s\) flow in the walls within the skin depth, sustaining the magnetic component of a resonant mode. These currents are confined near the surface and can be expressed as $\vec{J}_s = \hat{t} \times \vec{H}$, where \(\hat{t}\) is the unit vector tangent to the surface and \(\vec{H}\) is the magnetic field at the conductor boundary. Any structural discontinuity causes both the current  $\vec{J}_s$ to concentrate near the edges of the break and the resistance $R_s$ at the break.
%This leads to elevated local current densities \(|\vec{J}_s|^2\) in these regions, increasing resistive losses.
As a result, the power loss per RF cycle rises (thereby degrading $Q$-factor) as

\[
P_{\text{loss}} \propto R_s \int |\vec{J}_s|^2 \, dA,
\]
where the integral is over the cavity’s internal surface area \(A\). The slice orientation governs how severely these currents are disrupted. Latitudinal slicing tends to preserve the dominant azimuthal current paths for both modes unperturbed. Conversely, longitudinal slicing intersects current flowing regions, causing %greater breakage of \(\vec{J}_s\) pathways, increasing
power loss per RF cycle, therefore reducing $Q$ from $\sim$1,400 to $\sim$300.

Note that both slice directions cause disruptions to the system. Latitudinal slicing predominantly disrupts the electric field structure, while longitudinal slicing disrupts the magnetic field and, hence, the current distribution.
%, latitudinal slicing does so in a less invasive manner.
These electric or magnetic field-induced current disruptions manifest as distinct impedance signatures on the Smith chart, appearing as capacitive (for latitudinal slicing) or inductive (for longitudinal slicing), respectively.

The coupling method further defines the impedance, as observed on the Smith chart, due to the ratio between electric and magnetic fields changing for each modes. For a TM mode, the impedance is \(Z = \frac{E}{H}\). By design, the fundamental mode is excited through coupling to the magnetic field (loop coupler), while the harmonic mode is excited through coupling to the electric field (E-probe). Inductive coupling thus appears at a lower impedance on the Smith chart (smaller $\frac{E}{H}$ ratio), whereas electrical coupling appears toward higher impedance on the Smith chart (larger $\frac{E}{H}$ ratio), as shown in Figs.~\ref{Smith Chart}, \ref{experimental smith} conceptually and experimentally. When considered alongside slice orientation, the coupling strategy enables a quadrant-based segmentation of the Smith chart as shown in Fig.~\ref{Smith Chart}. This mapping provides a rapid and intuitive diagnostics for mode identification and verification of critical coupling, as shown in Fig.~\ref{experimental smith}.

\subsection{Material Factors}

The $Q$ factor of a cavity relates the stored energy $U$ to $P_{\text{loss}}$ as
\begin{equation}
Q_0 = \frac{\omega_0 U}{P_{\text{loss}}}.
\label{qu}
\end{equation}
%For the 3D-printed RF cavities, the dominant energy loss mechanisms are resistive losses and leaking RF signals. This occurs when the oscillating electromagnetic fields induce currents in the cavity walls or when the conductive material is insufficient to contain the RF signal. 
%The skin depth, $\delta$, is given by:
%\[
%\delta = \sqrt{\frac{2}{\mu \sigma %\omega}}
%\]
%where $\mu$ is the permeability of the material, $\sigma$ is its electrical conductivity, and $\omega$ is the angular frequency of the oscillating field.
%The skin depth is inversely proportional to the frequency, which means at higher frequencies, the skin depth decreases. This causes the surface currents to be confined to a thinner layer of the material. This effect is particularly relevant in RF cavities, where the current is confined to the internal surfaces of the cavity. As the skin depth shrinks (i.e the frequency increases), the conductivity of the material becomes even more critical because any deficiencies or inconsistencies in the conductive layer can lead to resistive losses.
In the presented plastic-based cavities, the conductive coating applied on the inner surfaces determines the conductivity and losses. Losses arise from surface roughness, metal film porosity and physical thickness that must be larger compared to the frequency-dependent skin depth $\delta = \sqrt{\frac{2}{\mu \sigma \omega_0}}$. %If the coating is not uniformly applied, areas of higher resistance form, creating hotspots for energy dissipation. These variations contribute to a reduction in the cavity’s ability to support electromagnetic fields efficiently, leading to a decrease in the field intensity and higher losses.
%Another factor that exacerbates losses is the potential leaking of the RF signal. This occurs when the cavity is not completely sealed or when there is a disruption in the conductive surface.

Silver and copper paints can have two to three orders of magnitude lower conductivity as compared to the bulk silver or copper \cite{copper-ink}, thereby requiring more than an order of magnitude thicker coatings, which is hard to control when painted manually. Thus, in addition to having higher $R_s$ (due to porosity, chemical solution residues, and surface roughness), painted coatings could also be RF leaky. Thin coatings based on vapor deposition are preferred as their conductivity is expected to be closer to that of bulk copper if thickness-uniform deposition is simultaneously possible in such a deposition process. From the comparison presented in Table~\ref{T:1}, improper coating and slicing can quadruple and double the power loss per RF cycle, respectively.

\section{Conclusions and Outlook}\label{Consclusion}

A series of four 3D-printed dual-mode (TM$_{010}$ and TM$_{011}$) cavities were designed, fabricated and tested. Three out of four cavities successfully passed baseline microwave measurements in that they 1) exhibited the fundamental and harmonic fields distribution; 2) resonated near their designed resonant frequencies; 3) could be critically coupled to 50 $\Omega$ line.

At the same time, there were critical lessons learned:

\noindent 1) Even though complex geometries can be printed by the modern 3D printers, one element can overcast another one, thereby making sputtering of the conductive coating challenging or even impossible. This led to one of four printed cavities to fail;

\noindent 2) Slicing is the most critical decision when designing a cavity. It was found that the best results (highest $Q$-factors and stonger mode fields) were obtained for the designs in which there were no continuity interruptions to the surface currents supporting particular mode magnetic field distribution. As highlighted in Table~\ref{T:1}, in the presented case of designing for TM$_{010}$ and TM$_{011}$ modes, the latitudinal slicing showed, by far, the best cavity performance.

\noindent 3) Lastly, it is highly desired to have a uniform and continuous conductive coating (about 5 skin depths) on the inner cavity surfaces. It is desired to have a pure material deposition (sputtering or electroplating plating) and not water or organic-based paints as conductivity of pure materials coatings is higher and, especially sputtering, are not intrinsically porous or rough as long as the plastic's surface is not porous or rough.

%retain The study revealed that the 3D-printed cavities performed adequately in terms of replicating the mode structures, with the longitudinally sliced, painted cavity yielding the best results. The lower electric field intensity and reduced Q-factors in the 3D-printed cavities can be attributed to several different factors such as; the inherent lower conductivity of the 3D-printed material, potential inconsistencies in the conductive coating and slice location. However, these differences were minimal in the context of the significant benefits that 3D printing provides, such as reduced production time and cost.

%Due to the surface current interactions, the 3D-printed cavities showed lower electric field intensity and Q-factor measurements compared to the copper cavity. Highlighting that the 3D printing can serve as an effective method for RF cavity fabrication, particularly in the prototyping phase.

In summary, based on best practices described here, 3D printing of RF cavities appears to be a viable and cost-effective approach for rapid prototyping or light weight cavity fabrication (likely for low and medium power applications).
%While further optimization of the printing process and coating techniques is critical to further improve conductivity and performance, the 3D-printed cavities have significant potential for future high-frequency applications, offering a cost-effective and efficient solution for cavity development and testing.

\section{Acknowledgments}\label{Acknowledgments}

The work was supported by the U.S. Department of Energy Office of Science, High Energy Physics under Cooperative Agreement Award No. DE-SC0018362.

\bibliography{references}

\appendix
\addcontentsline{toc}{section}{Appendices}
\renewcommand{\thesubsection}{\Alph{subsection}}

\end{document}